\documentclass[journal]{IEEEtran}

\usepackage[dvipsnames]{xcolor}
\usepackage{tikz}
\usepackage{hyperref}
\usepackage{multirow}
\usepackage{cite}
\usepackage{amsmath}

\ifCLASSOPTIONcompsoc
  \usepackage[caption=false,font=normalsize,labelfont=sf,textfont=sf]{subfig}
\else
  \usepackage[caption=false,font=footnotesize]{subfig}
\fi

\begin{document}

\title{Frequency dynamics of the Northern European AC/DC power system: a look-ahead study}
\author{Matas~Dijokas,~\IEEEmembership{Student Member,~IEEE,}
        Danilo~Obradovi\'{c},~\IEEEmembership{Student Member,~IEEE,}
        Georgios~Misyris,~\IEEEmembership{Student Member,~IEEE,}
        Tilman~Weckesser,~\IEEEmembership{Senior Member,~IEEE,} and
        Thierry~Van~Cutsem,~\IEEEmembership{Fellow,~IEEE}
\thanks{M. Dijokas and G. Misyris are with Technical University of Denmark, Department of Electrical Engineering, Kgs. Lyngby, Denmark (emails: {\mbox{matdij}, gmisy}@elektro.dtu.dk).  D. Obradovi\'{c} is with KTH Royal Institute of Technology, Stockholm, Sweden (e-mail: daniloo@kth.se). T. Weckesser is an independent reseacher (tilman.weckesser.dk@ieee.org).  T.~Van Cutsem was with the Fund for Scientific Research (FNRS) at the Dept. of Electrical Engineering and Computer Science, University of Li\`{e}ge, Belgium (e-mail: t.vancutsem@uliege.be).}  
\thanks{This work is supported by the multiDC project, funded by Innovation Fund Denmark, Grant Agreement No. 6154-00020B.}}

\maketitle

\begin{abstract}
A large share of renewable energy sources integrated into the national grids and an increased interconnection capacity with asynchronous networks, are the main contributors in reducing the kinetic energy storage in the Nordic Power System. Challenges arise to operate system after a loss of generation and to minimize the offshore systems interaction with onshore grids. To assess the associated challenges, a novel dynamic model was developed under the phasor approximation to represent the future Northern European Power System, including High Voltage Direct Current (HVDC) links and future offshore energy islands in the North Sea. First, the future frequency response is provided for two large disturbances to highlight the benefit of the developed model and point out potential future frequency issues. Consequently, further actions are investigated to better utilize the existing frequency containment reserves or to partially replace them using emergency droop-based power control of HVDC links. Lastly, the offshore grid interaction and frequency support to the Nordic network is investigated, and the dynamic response is compared for zero- and low-inertia designs of the offshore energy islands. The simulations were performed using industrial software, and the associated material is made publicly available.
\end{abstract}


\IEEEpeerreviewmaketitle

\section{Introduction}

\IEEEPARstart{T}{he} continuous drive to reduce carbon emissions and meet the climate agreement goals is reshaping power systems all over the world. Amongst many factors, the large-scale integration of inverter-based generation and the increasing interconnection capacity, contribute the most in phasing out conventional power plants, thereby decreasing power system kinetic energy storage. With lower inertia, the frequency response to power imbalances worsens, challenging Transmission System Operators (TSOs) to securely operate power systems \cite{OP2}. The problem has already been observed in island systems, e.g. Ireland or Australia \cite{EirGrid}, \cite{AEMO}, and it is progressively affecting larger interconnected systems as well, such as Great Britain or the Nordic countries (Norway, Sweden, Finland, and Denmark) \cite{GBlow}, \cite{CONPS}.   

\begin{figure}[t]
\centering
\includegraphics[width=.48\textwidth]{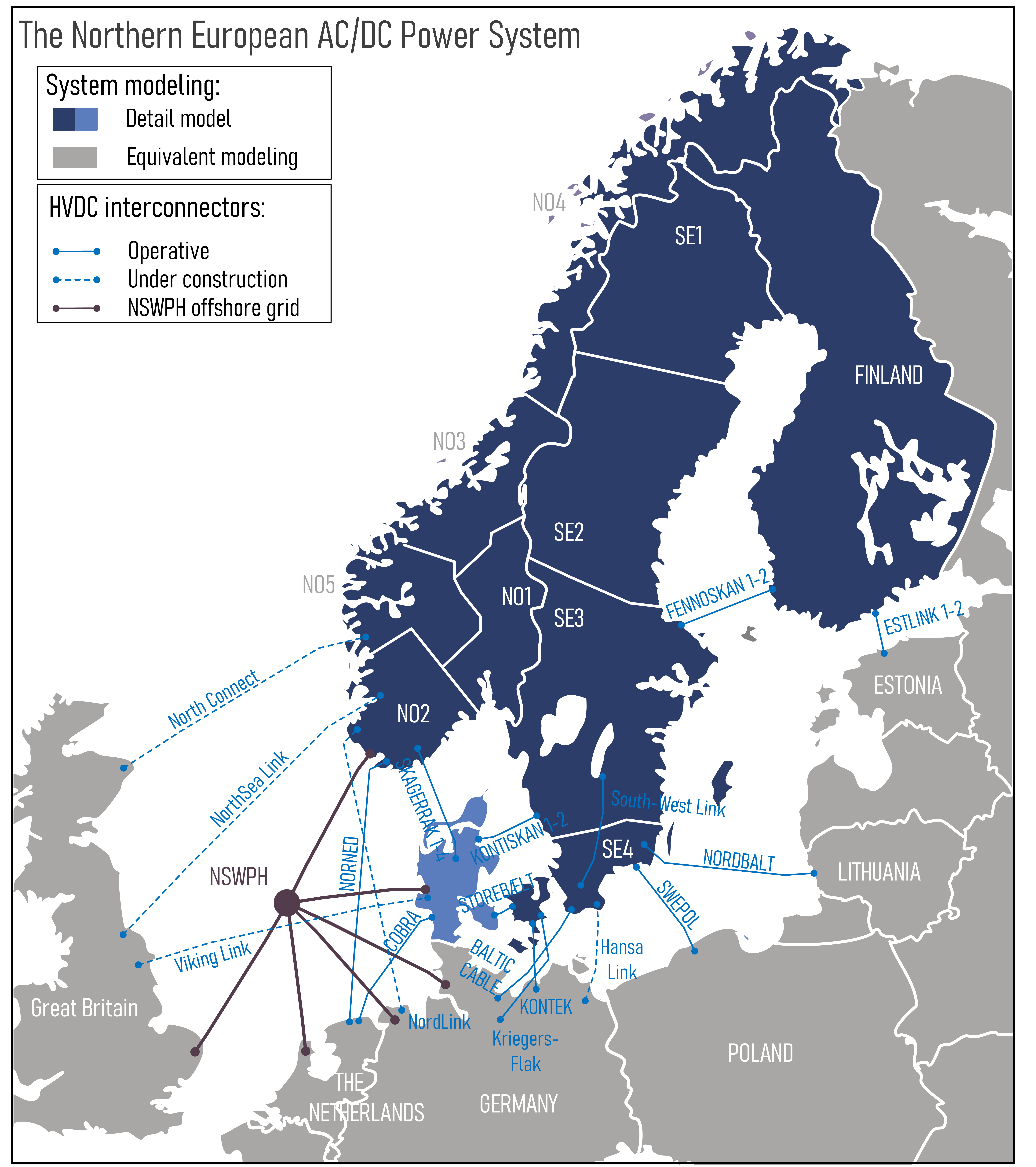}
\caption{Northern Europe, North Sea Wind Power Hub and HVDC interconnectors. The model presented in the paper is shown in blue. The dark blue part corresponds to the Nordic Power System. Numbers appearing in line names correspond to multiple HVDC line in parallel.}
\label{fig:Map}
\end{figure}

The Nordic Power System (NPS), corresponding to the dark blue color in Fig.~\ref{fig:Map}, has experienced a progressive growth of wind generation with a cumulative power almost doubled over the past 10 years \cite{IRENA} and the forecast shows that capacity is expected to double again by 2030 \cite{WindEurope}. Moreover, global policies to reinforce the interconnected European grid and increase generation from renewable energy sources \cite{EU_policy} are already taking effect with additional 4.9~GW of new interconnection capacity scheduled to be operational by 2025 (see Fig.~\ref{fig:Map}) \cite{CONPS} and the commitment to build a cluster of islands harvesting up to 36~GW of wind energy in the North Sea \cite{NSWPH_policy}. As a result, by 2040, the total kinetic energy of synchronous generators in the NPS may drop below 125~GWs for more than 2100 hours per year threatening secure and reliable operation of the system \cite{Svk_Marketanalysis}. Several reports have already addressed the decreasing kinetic energy issue in the NPS \cite{CONPS}, \cite{FSI2}, \cite{DI2}, with system frequency falling below the allowed lower value of 49.0~Hz, following the dimensioning incident.
It has been also shown that the current Frequency Containment Reserve (FCR) (or primary frequency control reserve) requirements might not be sufficient to contain frequency deviations within the given limits, and other frequency control means are needed \cite{FCR_requirements}. A number of solutions have been already proposed in the literature \cite{FSI2}, \cite{OP1} -\cite{ADC}, amongst which the most promising is Emergency Power Control (EPC) of HVDC interconnectors. 

Compared to other solutions such as virtual inertia schemes or down-regulation of critical units \cite{textbook}, \cite{CA_EPC}, HVDC EPC takes advantage of existing infrastructure (AC/DC converters with local frequency measurements) and does not require substantial regulatory changes, such as new grid codes and/or market rules. Given the large number of HVDC links connecting the NPS to asynchronous areas and the future availability of additional links connecting the NPS to the planned North Sea Wind Power Hub (NSWPH) (see Fig. 1), the support from HVDC links is a very promising solution to counteract frequency deviations and allow lower FCR reserves. The dynamic performance of HVDC EPC is presented in \cite{NAPS} where it is shown to supplement FCR response and various approaches are compared in terms of Instantaneous Frequency Deviation (IFD) improvement and reserve capacity usage. Reference \cite{DAN_PSS} shows analytically how EPC control with appropriate frequency droop gains can also improve both synchronizing and damping torque components of synchronous machines. As regards to the future NSWPH, Ref.~\cite{Cigre_NSWPH} presents a scaled-down concept of a single energy island and assesses the stability of this AC system asynchronous with onshore grids. However, in all the aforementioned publications, the frequency responses are obtained from single-mass equivalent models, thereby neglecting individual, local impacts of AC/DC converters and other system components such as loads and generators.

On that account, this paper aims at reporting on the future NPS frequency response in the context of strategic system changes foreseen for the 2020 - 2030 period, and how HVDC support can complement and/or replace FCR during low kinetic energy hours. Furthermore, the operation of the NSWPH is analyzed, in particular its impact on the onshore networks when its HVDC links are equipped with the aforementioned HVDC EPC control. The whole work has been based on a realistic multi-machine model under the phasor approximation. The corresponding material can be publicly accessed and used by other researchers. In detail, the contributions of this paper can be summarized as follows:
\begin{itemize}
    \item A model of the NPS primarily intended to study frequency dynamics. This involves individual (operational and under construction) point-to-point HVDC links with generic Line Commutated Converter (LCC) and Voltage Source Converter (VSC) models, appropriate dependency of loads to local voltage and synchronous generator response obeying the current FCR dynamic requirements;
    \item An improved frequency support by HVDC links, based on frequency droop control. The analysis shows how they can be exploited to limit IFD in the NPS and which aspects have to be taken into account for FCR replacement with HVDC EPC;
    \item A modular energy island concept whose purpose is to collect large amounts of offshore wind energy and serve as a hub for power exchanges between the connected onshore grids. The control encompasses either low- or zero-inertia topologies. The island is also participating in NPS frequency support by exploiting the onshore frequency error signals to control the offshore AC/DC converters.
\end{itemize}

The rest of the paper is organized as follows. An overview of the NPS and its model is given in Section II. Sections III and IV present the salient features of the system: NSWPH and HVDC EPC, respectively. Dynamic simulation results are provided and discussed in Section V. Concluding remarks are offered in Section VI.

\section{The Northern European system and its model}

\subsection{Frequency control in the NPS}

The NPS makes up one synchronous area with a nominal frequency of 50 Hz. It is operated by four TSOs. The total load ranges from 25 to 70~GW, while the total kinetic energy of synchronous generators varies in between 125 and 240~GWs at the moment. Currently, the grid is connected to the rest of Northern Europe through 18 point-to-point HVDC links and five more are being developed at the moment (see Fig. \ref{fig:Map}) \cite{FSI2}.

Frequency regulation resorts to the following main services: FCR for Normal operation (FCR-N), FCR for Disturbances (FCR-D) and Frequency Restoration Reserve (FRR). In normal system operation the frequency deviations are limited to $\pm$100~mHz and \mbox{FCR-N} is deployed to keep the frequency inside that standard band. In case of a larger power imbalance, when frequency drops below 49.9~Hz, FCR-D is activated. FRR is used to restore the frequency back to nominal value. The system is designed to operate with a system stiffness \cite{Kundur} of at least 3625 MW/Hz with the objective of limiting the steady-state frequency deviation to $\pm$ 500~mHz \cite{NPS:BP}.

In the NPS, the maximum IFD must be smaller than $\pm$1~Hz. Under-frequency load shedding is activated at 48.8~Hz. Although its activation is deemed unlikely (only for a large disturbance, low inertia, and poor FCR performance), it would lead to a heavy compensation cost. Therefore, the NPS TSOs have set the lowest allowed frequency to 49.0~Hz and they have accordingly determined actions to decrease the risk of load shedding activation.

Currently, the dimensioning incident is the outage of Oskarshamn-3, a 1450-MW nuclear power plant in Sweden.

\subsection{Overview of the model}

The developed model, corresponding to the blue areas in Fig.~\ref{fig:Map}, includes the following:
\begin{itemize}
\item Norway (NO), Sweden (SE) and Finland (FI) represented only with their extra-high-voltage transmission network (300 and 400~kV), which include large number of long transmission lines;
\item The eastern part of Denmark (DK2), which belongs to the same synchronous (NPS) zone. Owing to the large number of HVDC links connected to Denmark, it was decided to represent the Danish grid in greater detail, down to lower voltage levels (400, 165 and 150~kV). This involves much shorter lines/cables and a comparatively large number of substations;
\item The western part of Denmark (DK1), which is synchronous with Continental Europe (CE). For the same reason as above, that part of the Danish grid involves lower voltage levels.
\end{itemize}
The main characteristics of the model are provided in Table~\ref{tab:NSS_overview}.

At the other end of HVDC links, the neighbouring synchronous zones, in particular Great Britain (GB) and CE, are represented with single-mass equivalents and simplified frequency control.
 
The model takes into account the strategic changes identified in \cite{CONPS}: increased wind power production, decommissioning of nuclear power plants, internal grid reinforcements and new interconnections. 
 
The focus of the study is on frequency and long-term dynamics (lasting up to - say - one minute after a disturbance), which led to represent slow controls such as Load Tap Changers and shunt compensation switching. The model has been developed under the phasor approximation \cite{Kundur}. The DigSilent PowerFactory 2018 software has been used to that purpose \cite{PF2018}. Since all components are represented by generic models and no proprietary information is involved, the data have been made publicly available and can be accessed at \cite{GitHub}. 
  
\begin{table}[]
\centering
\caption{Summary of the main model characteristics}
\label{tab:NSS_overview}
\begin{tabular}{|c|c|c|c|c|c|}
\hline
\multicolumn{1}{|l|}{}    & \multicolumn{4}{c|}{\textbf{Nordic Power System}}      & \multirow{2}{*}{\textbf{DK1}} \\ \cline{2-5}
                          & \textbf{NO} & \textbf{SE} & \textbf{FI} & \textbf{DK2} &                               \\ \hline
\# of buses               & 45          & 30          & 13          & 140          & 171                           \\
\# of transmission lines  & 35          & 27          & 5           & 174          & 165                           \\
\# of generators          & 16          & 9           & 5           & 17           & 29                            \\
\# of WPPs                & 4           & 6           & 2           & 17           & 35                            \\
\# of loads (neg. loads)  & 15          & 13          & 9           & 139(73)      & 135(99)                       \\
\# of compensation units  & 2           & 3           & -           & 21           & 62                            \\ \hline
Total generation {[}MW{]} & \multicolumn{4}{c|}{43723.32}                          & 2112.8                        \\
Total load {[}MW{]}       & \multicolumn{4}{c|}{43531.2}                           & 2936.3                        \\ \hline
\# of HVDC lines          & \multicolumn{5}{c|}{24}                                                                \\ \hline
\end{tabular}
\end{table}

The generation mix is dominated by hydro and thermal power plants. In total, 76 generation units with nominal power above 40 MVA are represented behind their step-up transformers. All synchronous machines have a $5^{th}$ or $6^{th}$ order model (equivalent to model 2.1 and 2.2 according to \cite{IEEEgen}), including saturation effects, 
supplemented with automatic voltage regulator and power system stabilizer models. The latter were tuned to ensure a sufficient level of rotor angle stability. The generators below 40~MVA were represented as negative loads.

Ten generators participate in FCR-D and these are equipped with hydro turbine/governor models. The corresponding parameters were selected in realistic ranges of values, and using the methodology in \cite{NAPS} to satisfy the existing requirements for FCR-D.

The consumption is represented as aggregated loads at specific voltage levels. The dependency of load power to voltage is represented with a ZIP model \cite{Kundur}. Its parameters were adopted from  \cite{CigreLoad}, in which loads were aggregated by type in each country and the model was validated against post-event measurements. No frequency dependency of loads has been considered, which is a little pessimistic.

The model includes 24 HVDC links, of which 13 involve AC/DC converters of the LCC type and 11 of the VSC type. Each LCC-HVDC link is modeled as a two-port element containing a DC cable, a rectifier and an inverter. The converters and their controls have generic models according to \cite{Cigre_benchmark}. Each LCC converter station includes a transformer with load tap changer and switchable shunt capacitor banks. The transformer ratios and the shunt susceptances are adjusted by ``slow'', discrete controllers. Each VSC-HVDC link is modeled by a two-port element containing a DC cable, two two-level converters, DC capacitors and phase reactors. The converters are equipped with grid-following Synchronous Reference Frame (SRF) controls, as presented in \cite{Cigre_NSWPH}. The controllers were carefully tuned to exhibit adequate dynamics. For instance, the response time after a step change of active power reference is within 200~ms for LCC links and 100~ms for VSC links. Most of the HVDC links operate in active power control mode, while VSCs are set to regulate the voltage at their points of connection behind a phase reactor. 

Finally, the Wind Power Parks (WPP) are represented with grid-following SRF control including only the outer control loops (thus neglecting inner loops). The model does not consider any plant level control and assumes operation for maximum power point tracking.

\subsection{System operating point}

The majority of the power is provided by plants located along the Norwegian Coast and in Northern Sweden (see NO3, NO4, NO5, SE1, and SE2 zones in Fig.~\ref{fig:Map}). The power is transferred along a North-South axis to meet high load demand in densely populated areas (see NO1, NO2, SE3, SE4, and FI zones in Fig.~\ref{fig:Map}). The total load in NPS is 43.5~GW, of which 14.9~GW are covered by WPPs. The kinetic energy storage of the NPS system is around the low value of 125~GWs. The operating point was adjusted according to the commonly observed power flow orientations in HVDC interconnections and load centers, so that a plausible scenario for the future system is considered.

\section{The North Sea Wind Power Hub and its model} \label{sec:NSWPH}

The NSWPH considered in the study would be located as shown in Fig.~\ref{fig:offshore_topology}. Among many topologies, the AC hub emerges as a possible technical solution to accommodate and connect electrical equipment. The considered topology in this paper consists of three identical islands connected through 400-kV submarine AC cables. The Hub-and-Spoke design, currently considered by the involved TSOs \cite{NSWPH_Vision}, allows modular expansion and isolation of an island in disturbed operation conditions.

\definecolor{green_top}{rgb}{0.3294,0.5098,0.2078}%
\definecolor{blue_map}{rgb}{0.2863,0.3294,0.4471}
\definecolor{grey_map}{rgb}{0.3490,0.3490,0.3490}
\definecolor{dred}{rgb}{0.3254,0.23529,0.29803}
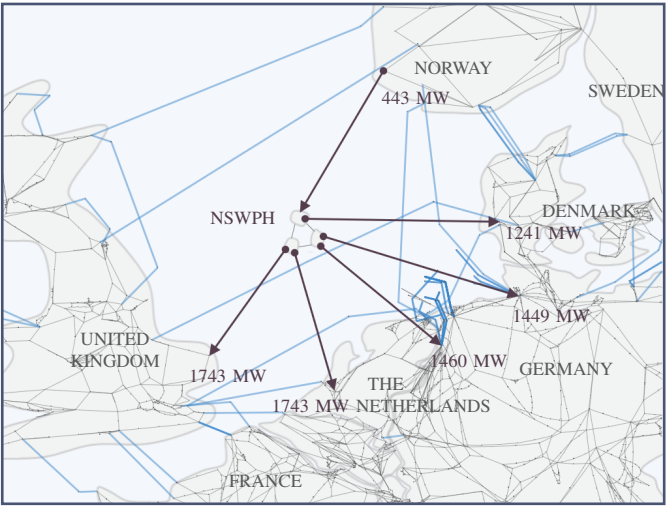
\begin{figure}[h]
    \centering
    \begin{tikzpicture}
        \node[inner sep=0pt, anchor = south west] (network) at (0,0) {\includegraphics[width=.485\textwidth]{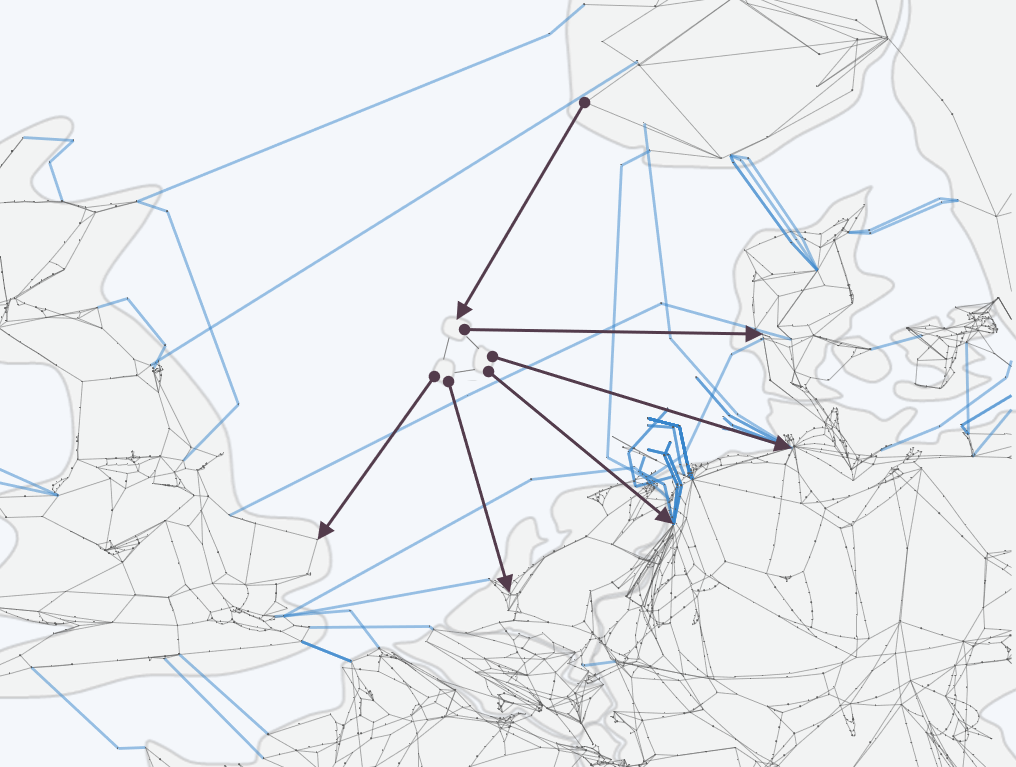}};
        \fill[fill=none, draw=blue_map, line width=1] (0,0) rectangle (8.8,6.65);
        \node[] at (3.2,3.8) {\scriptsize \textcolor{dred}{NSWPH}};
        \node[] at (1.5,2.2) {\scriptsize \textcolor{grey_map}{UNITED}};
        \node[] at (1.5,1.9) {\scriptsize \textcolor{grey_map}{KINGDOM}};
        \node[] at (6.0,5.8) {\scriptsize \textcolor{grey_map}{NORWAY}};
        \node[] at (7.8,3.9) {\scriptsize \textcolor{grey_map}{DENMARK}};
        \node[] at (7.5,1.8) {\scriptsize \textcolor{grey_map}{GERMANY}};
        \node[] at (3.5,0.3) {\scriptsize \textcolor{grey_map}{FRANCE}};
        \node[] at (5.1,1.6) {\scriptsize \textcolor{grey_map}{THE}};
        \node[] at (5.6,1.3) {\scriptsize \textcolor{grey_map}{NETHERLANDS}};
        \node[] at (8.3,5.5) {\scriptsize \textcolor{grey_map}{SWEDEN}};
        
        \node[] at (3.0,1.7) {\scriptsize \textcolor{dred}{1743 MW}};
        \node[] at (4.1,1.3) {\scriptsize \textcolor{dred}{1743 MW}};
        \node[] at (6.2,1.9) {\scriptsize \textcolor{dred}{1460 MW}};
        \node[] at (7.3,2.5) {\scriptsize \textcolor{dred}{1449 MW}};
        \node[] at (7.2,3.6) {\scriptsize \textcolor{dred}{1241 MW}};
        \node[] at (5.5,5.4) {\scriptsize \textcolor{dred}{443 MW}};
    \end{tikzpicture}
    \caption{Overview of three-island NSWPH topology with point-to-point HVDC connections (in dark red) to the onshore grids. The 400-kV transmission lines in the onshore grids are shown with thin black lines.}
\label{fig:offshore_topology}
\vspace{-0.5em}
\end{figure}

\begin{figure}[h]
\centering
\includegraphics[width=0.9\columnwidth]{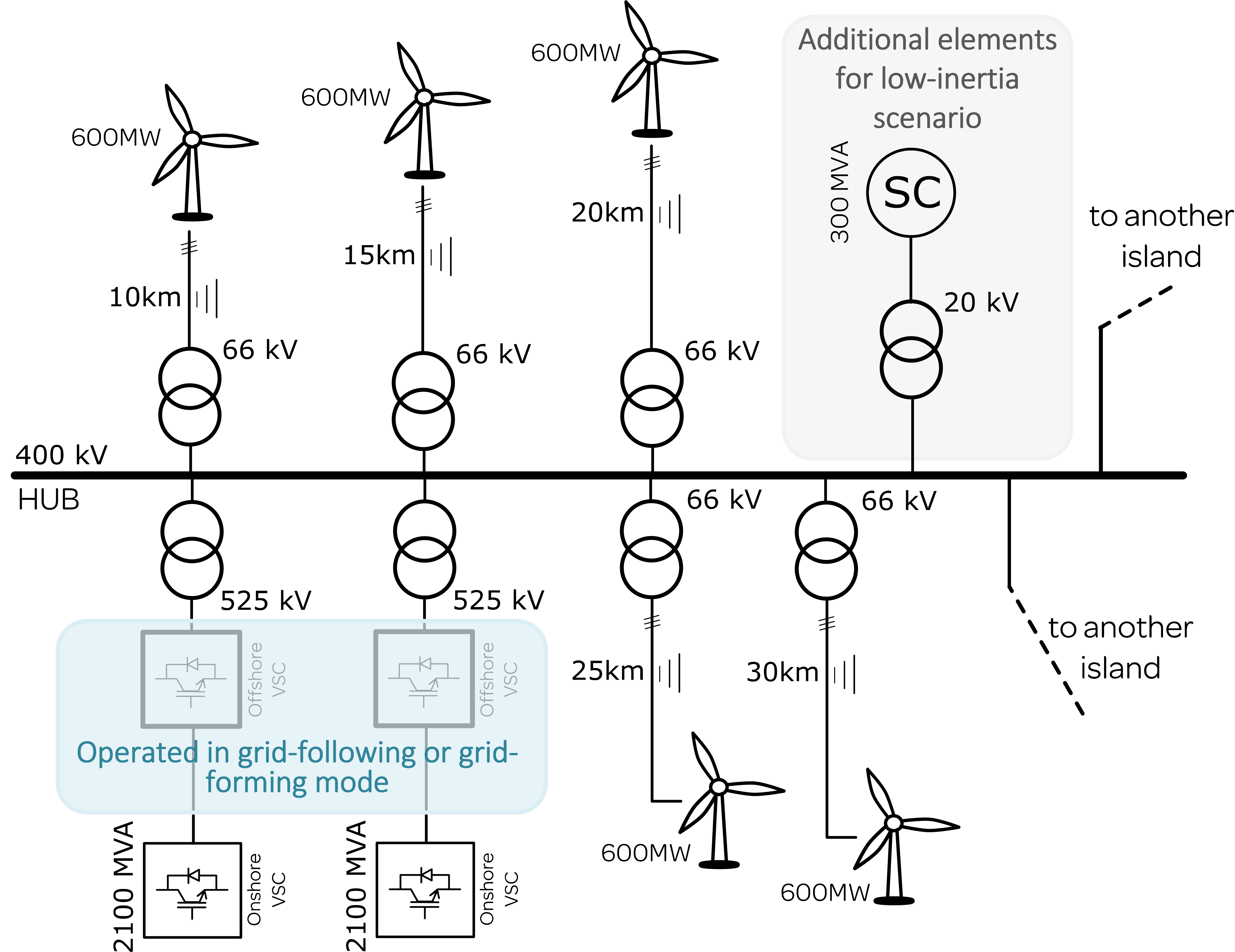}
\caption{AC grid topology in one of the NSWPH islands (low- and zero-inertia configurations)}
\label{fig:NSWPH_Topology}
\end{figure}

The total installed capacity of the offshore WPPs connected to the NSWPH is 9~GW, each island harvesting 3~GW of wind generation. As shown in Fig.~\ref{fig:NSWPH_Topology}, in the model, the WPPs connected to each island are lumped into five equivalent WPPs, connected to a 400-kV local hub through 66-kV cables. Each equivalent WPP collects the same power but the length of the 66-kV connection varies from one WPP to another to account for their positions around the island. Six 2-GW \mbox{$\pm$525-kV} VSC-HVDC links, two per island, connect the NSWPH to three onshore synchronous zones, namely GB, CE and NPS. The WPPs are operated in constant PQ mode with only the DC/AC converter dynamics represented; hence, they do not contribute to frequency or voltage control.

Two control configurations are available in the model, referred to as zero- and low-inertia.

\begin{figure} [!h]
    \centering
  \subfloat[\vspace{-3ex}\label{fig:Ng1}]{%
       \includegraphics[width=0.45\textwidth]{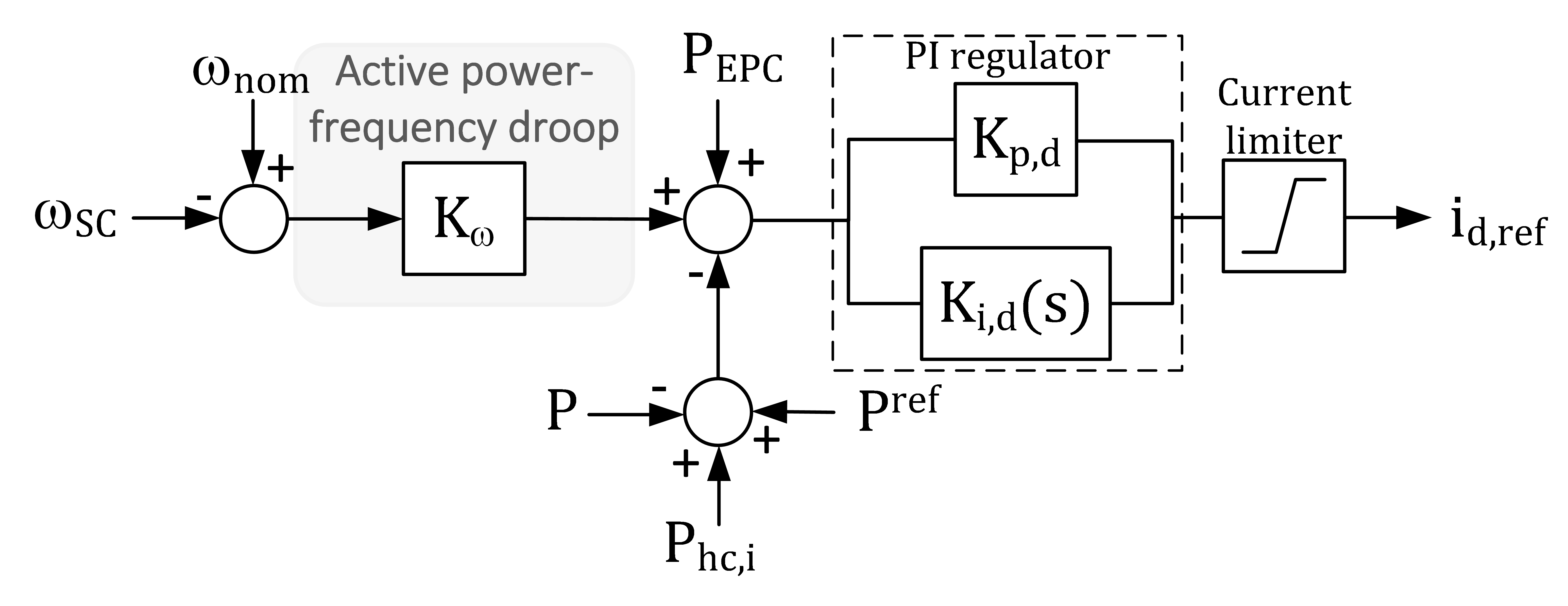}}
    \\
  \subfloat[\vspace{-3ex}\label{fig:Ng2}]{%
        \includegraphics[width=0.4\textwidth]{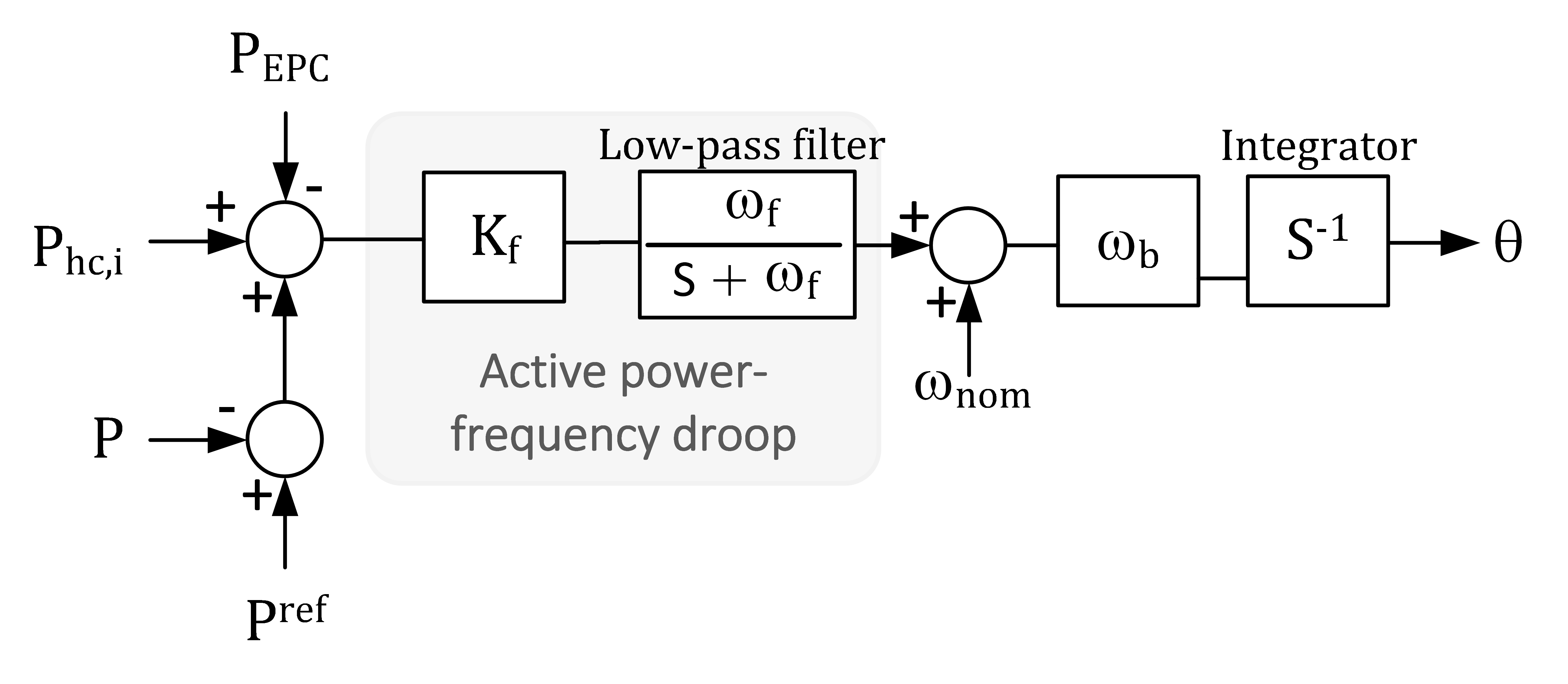}}
    \\
  \subfloat[\label{fig:Ng3}]{%
        \includegraphics[width=0.35\textwidth]{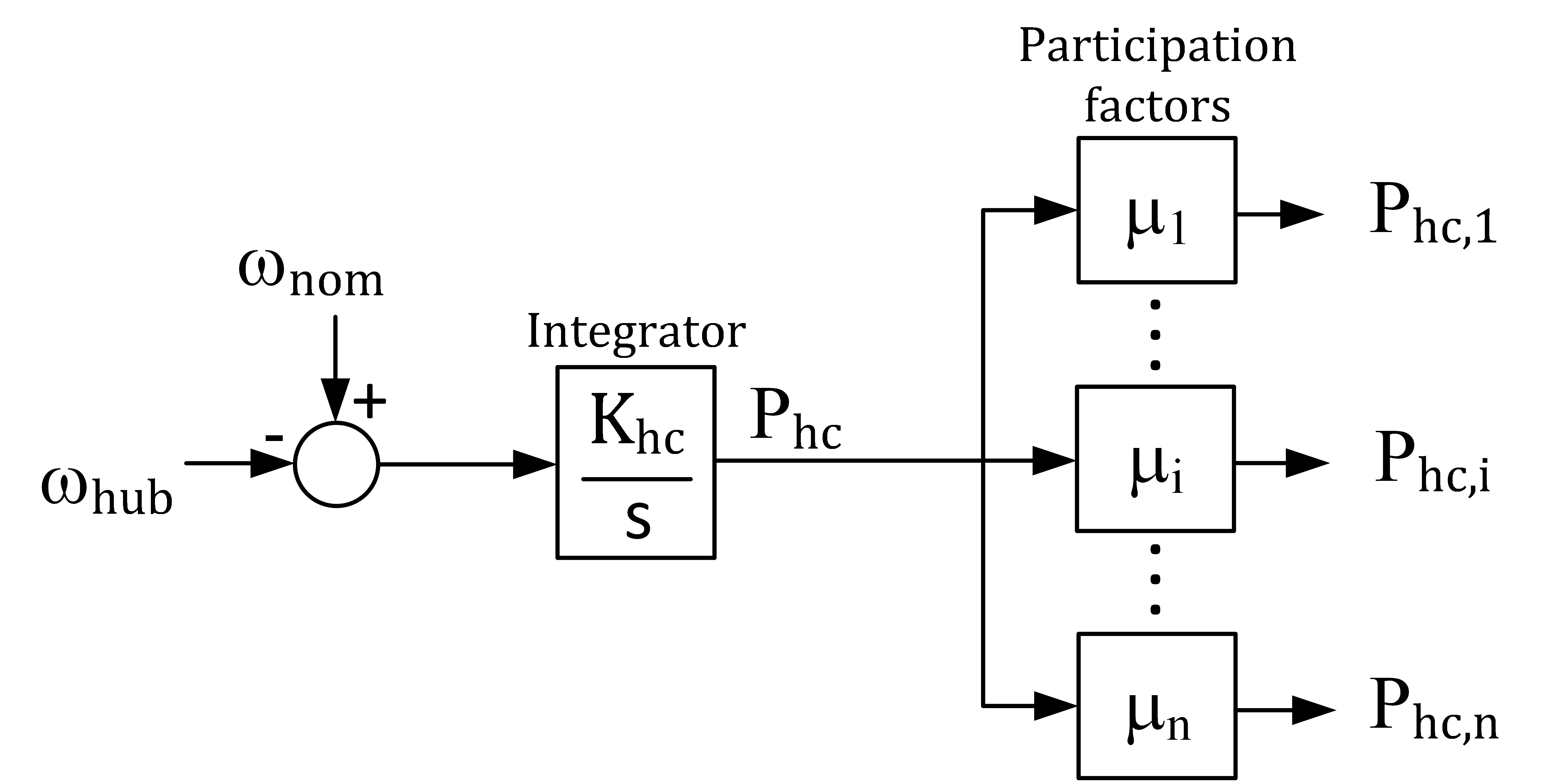}}
  \caption{Outer control loop block diagram of the offshore VSC converters. (a) grid-following control mode. (b) grid-forming control mode \cite{GM_GF}. (c) hub coordinator control.}
  \label{fig1} 
\end{figure}

In the low-inertia approach \cite{Cigre_NSWPH}, \cite{PSCC2020}, each AC island is equipped with a Synchronous Condenser (SC), as shown in Fig.~\ref{fig:NSWPH_Topology}. The frequency in the AC islands is set by the SC rotor speeds. Their kinetic energy storage allows smoothing the impact of offshore disturbances on onshore grids. The SCs, equipped with automatic voltage regulators, not only control the grid voltages but also provide a reference with which the VSCs synchronize. Hence, in this approach, each VSC can operate in grid-following mode, tracking the grid voltage phasor with a Phase-Locked Loop and adjusting the phasor of its injected current in accordance with the desired active and reactive powers. Through the outer control loop shown in Fig.~\ref{fig:Ng1}, each VSC imposes an active power-frequency droop. The frequency deviation is slowly corrected by a ``hub coordinator'' with integral action on the frequency error as shown in Fig.~\ref{fig:Ng3}. It can also redistribute the power changes among the various VSCs according to participation factors. The nominal power of each SC is 300~MVA, a value that preserves stability of the NSWPH grid after the outage of any of the three SCs. The inertia constant $H$ has been set to 2~s.

Zero-inertia refers to the absence of any rotating machine, i.e. it is a 100\% converter-based system \cite{Cigre_NSWPH}, \cite{SC}. Hence, the offshore VSCs operate in grid-forming mode. Each VSC imposes the magnitude and phase angle of its modulated AC voltage, behind its step-up transformer. The various VSCs synchronize with each other through their individual control loops adjusting the frequency of each modulated voltage based on the difference between the measured active power and its set-point, according to some droop as shown in Fig.~\ref{fig:Ng2}. The zero-inertia configuration allows keeping the NSWPH frequency close to its nominal value (50~Hz) and saves the footprint of the SCs. On the other hand, offshore power imbalances are quickly propagated to the DC voltages of the HVDC links and to the onshore grids. After a large disturbance any over-current in a grid-forming VSC is promptly corrected through the virtual impedance controller to avoid damage \cite{VI_control}.

\vspace{-1ex}

The links connected to GB and CE export the power from the hub. Fig.~\ref{fig:offshore_topology} shows the power flows at the initial operating point. They would correspond to a common situation where Norway exports power \cite{AT_NSWPH}, in this case 443~MW flowing from Norway to the NSWPH.

\section{Emergency Power Control of HVDC links}

\begin{figure}[h]
\centering
\includegraphics[width=.38\textwidth]{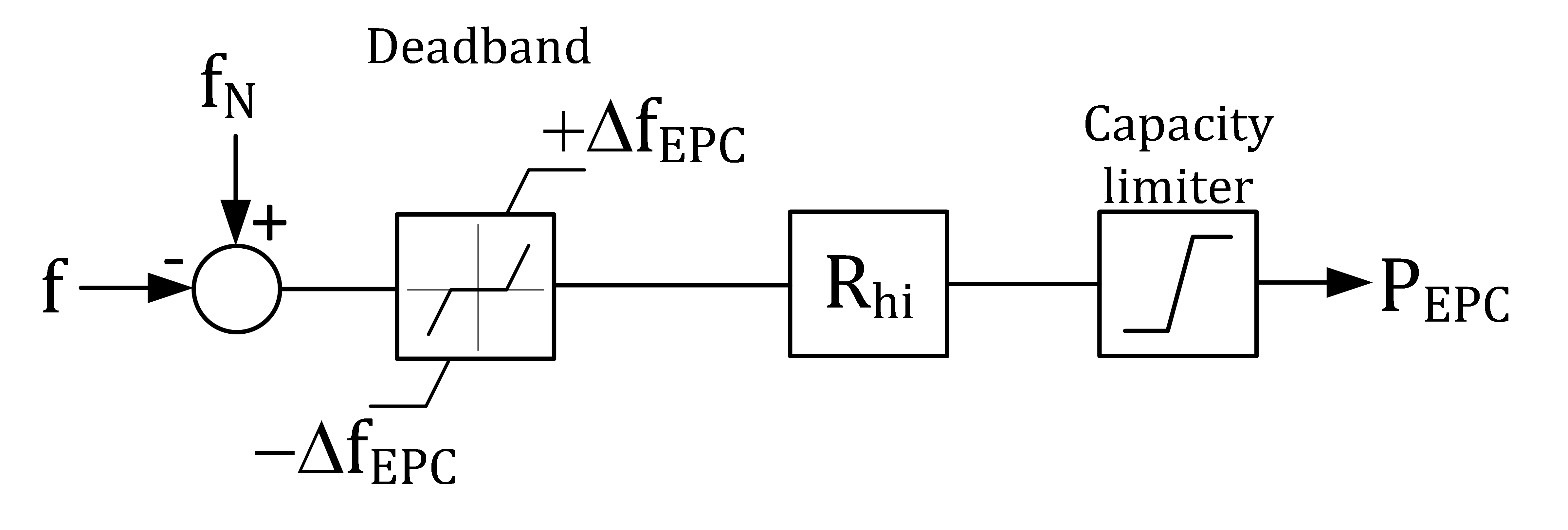}
\caption{Droop control for HVDC EPC (sign are adjusted for HVDC power import to NPS case)}
\label{fig:EPC1}
\end{figure}

The aim of the HVDC EPC is to aid the conventional units involved in FCR-D to limit the IFD, namely to keep frequency above 49.0~Hz after the dimensioning incident. This emergency control relies on a corrective signal sent to the active power controllers of designated AC/DC converters in order to promptly increase or decrease the active power injection into the NPS. EPC relies on a closed-loop design as shown in Fig.~\ref{fig:EPC1}, in which the active power correction is proportional to the frequency deviation obtained from local measurement. More precisely, EPC is activated when the input frequency signal $f$ reaches the triggering threshold $f_{TFL}$, after which the active power reference is adjusted by $P_{EPC}$ proportionally to the frequency error $f_{TFL}-f$. When the frequency signal regains a value above $f_{TFL}$ (i.e. $f_{TFL}-f \leq 0$), the active power reference $P_{EPC}$ is reset to zero, disabling the EPC correction of the converters. 

\begin{figure}[h]
\centering
\includegraphics[width=.48\textwidth]{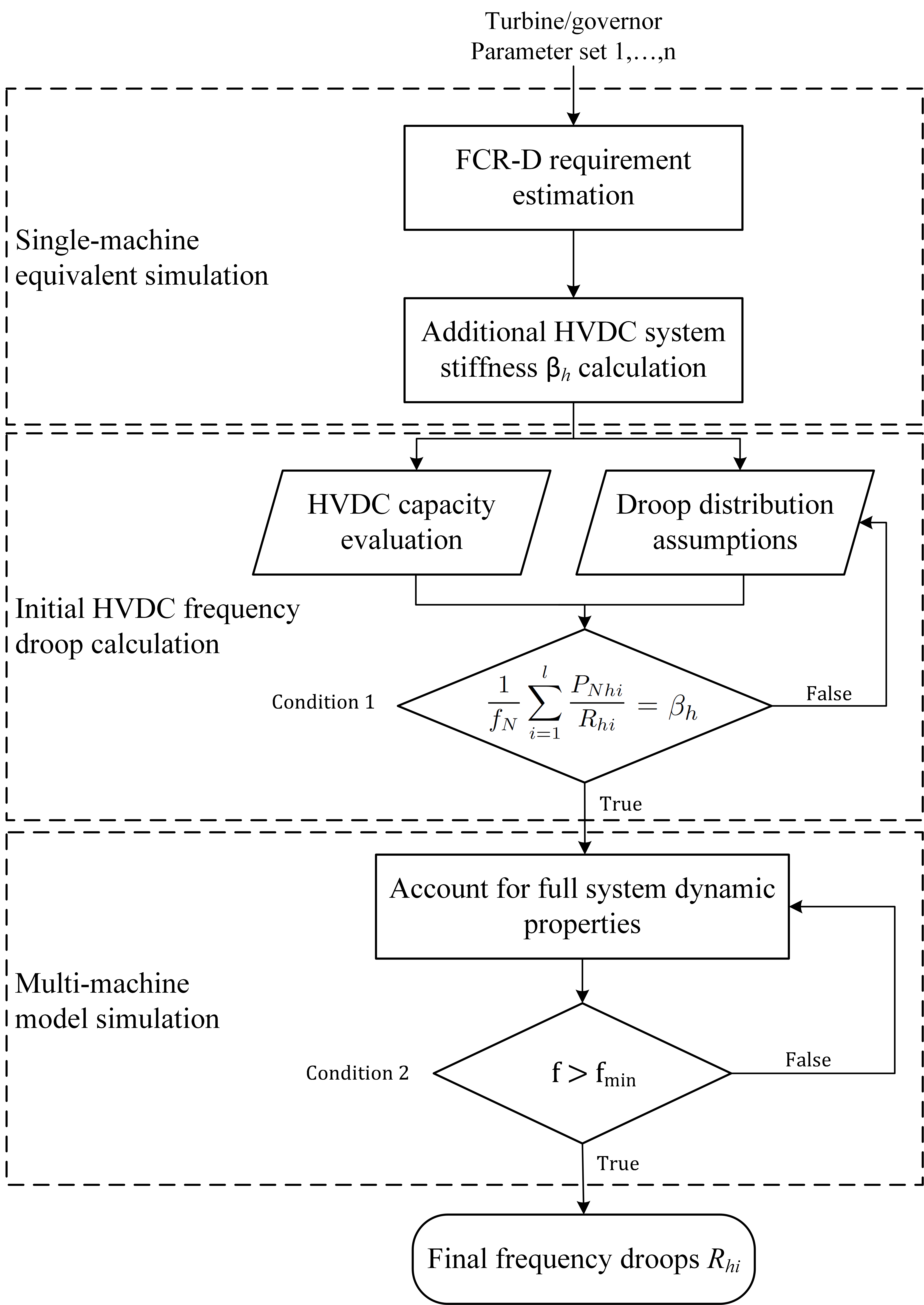}
\caption{A flowchart to determine HVDC EPC frequency droop in pu for individual HVDC links using single- and multi-machine models}
\label{fig:algorithm}
\end{figure}

Out of the 24 HVDC links present in the model, 18 connect the NPS with asynchronous areas and one with the NSWPH. They are assumed to be all equipped with EPC. The activation time delays are assumed negligible, since local measurements are used. The activation threshold $f_{TFL}$ has been set to 49.6~Hz, while the more conservative value of 49.9~Hz was used in \cite{NAPS}. The active power correction in each HVDC link is limited according to its operating point in order not exceed the power capability of the converters. The droop between $f_{TFL}-f$ and $P_{EPC}$ has been determined using the equivalent droop method detailed in \cite{NAPS}. The corresponding flow chart is given in Fig. \ref{fig:algorithm}.  First, for a realistic set of turbine/speed governor parameters satisfying FCR-D requirements and for the dimensioning incident, the HVDCs system stiffness ($\beta_{h}$) is identified for a given total kinetic energy level in NPS, using the single-machine equivalent model in \cite{NAPS}. Equal droops in per unit are assumed for all HVDC links. Next, an iterative procedure adjusts those individual frequency droops ($R_{hi}; i=1,...,n$) taking into account the available HVDC capacities, in order to keep the equivalent droop at NPS level equal to $\beta_{h}$. Finally, the individual droop values are adjusted based on simulations of the detailed, multi-machine system to account for the impact of other dynamic system properties such as power system stabilizers and the dependency of load power to voltage.

EPC is a supplementary frequency control aimed at complementing FCR-D, but its use can be extended to partly replacing conventional FCR-D procured within the synchronous zone. The share of frequency control between FCR-D and EPC is further discussed in the Appendix. When replacing \mbox{FCR-D} with EPC, the HVDCs system stiffness ($\beta_{h}$) can be calculated from (\ref{eq:Fss}) in Appendix which yields the minimum system stiffness provided by HVDCs to ensure a final steady-state frequency above 49.5~Hz. The latter is used as an input to calculate the HVDC droops ($R_{hi} (i=1,...,n)$) and the final frequency droop values, according to Fig.~\ref{fig:algorithm}. However, the "\textit{Condition 2}" used in multi-machine model simulations is changed to satisfy \mbox{steady-state} frequency deviation constraint ($\Delta f_{ss} >\Delta F_{ss,max}$).   

Clearly, the frequency nadir after a generator outage depends not only on the frequency droop values, but also on the settings of speed governors and HVDC outer control loops. The HVDC active power controller time response is tuned to be in the $100-200$~ms range, which is considerably faster compared to hydro plant speed governor responses. Thus, replacing FCR-D with EPC will yield higher frequency nadir values (or lower IFD) compared to a case when only generators are used for FCR-D.

\section{Simulation Results}

The simulation results presented in this section deal with the frequency dynamics of the future Northern European AC/DC system, including the NPS and the NSWPH, without and with EPC, respectively. The simulations involve large disturbances such as generator or HVDC interconnector outages. They illustrate various features of the model that may be of interest to other researchers.

In all simulations, the system is initially in steady-state with the NPS frequency at 49.9~Hz (the lowest value with FCR-D reserves not activated) and the NSWPH frequency at 50~Hz.

The contributions to system stiffness of generators ($\beta_g$) and HVDC links ($\beta_h$) are defined in the Appendix.

\subsection{NPS frequency dynamics with FCR-D from generators}\label{ch:res_Fnps}

Two disturbances differing by the size and the location are studied. The first one is the dimensioning incident outage of the Oskarshamn-3 \mbox{1450-MW} nuclear power plant. This is currently considered as the most critical incident in NPS. This generator is located in zone SE3 (see Fig.~\ref{fig:Map}) serving a relatively high load of 9~GW. The second incident is the loss of the Rossaga-G1 \mbox{1040}-MW generator. It is located in zone NO4. This zone serves a comparatively low load of 2~GW.

These two representative disturbances allow illustrating the impact on frequency response of voltage-dependent loads as well as proximity of the disturbance to load centers. Although these effects are sometimes disregarded in the analysis of frequency control, they may be significant for proper design of frequency reserves in low-inertia systems.

\begin{figure}[t!]
\hspace*{-3ex}\includegraphics[width=.555\textwidth]{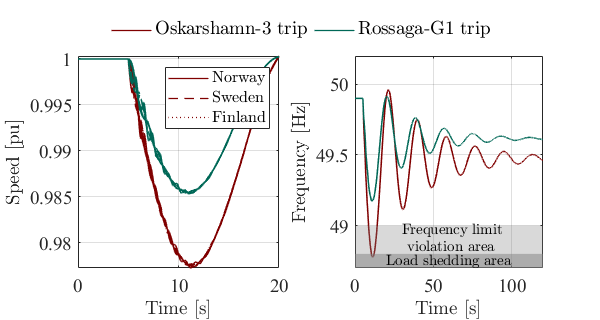}
\caption{NPS frequency response to two disturbances. Left plot: individual speeds of FCR-D units. Right plot: average frequency of \mbox{FCR-D} units}\label{fig:f_system}
\end{figure}

Fig. \ref{fig:f_system} presents the frequency responses to the aforementioned disturbances. The left plot shows the rotor speeds of individual machines involved in FCR-D, while the right plot shows the average frequency of these units over a longer time interval. The left figure shows, as expected, that both disturbances excite local modes of oscillation that can be observed on individual generators. Those oscillations are adequately damped by the PSS, so that after some 13 seconds, all generators follow the global (common) mode of oscillation. 

In response to the Oskarshamn-3 outage, the frequency nadir reaches a low value of 48.78~Hz, which violates the maximum IFD criterion (at 49.0~Hz). Even more, the nadir value is just at the limit of activating load shedding in the NPS (at 48.8~Hz). Finally, the system stiffness is not sufficient to limit the steady-state frequency deviation to the required value of 0.5~Hz after the dimensioning incident. Expectedly, the results are less severe for the Rossaga-G1 outage. The FCR-D units are able to keep the frequency within the desired range with a nadir at 49.17~Hz and a final frequency deviation of 0.4~Hz. 

\begin{figure}[t!]
\hspace*{-3ex}\includegraphics[width=.555\textwidth]{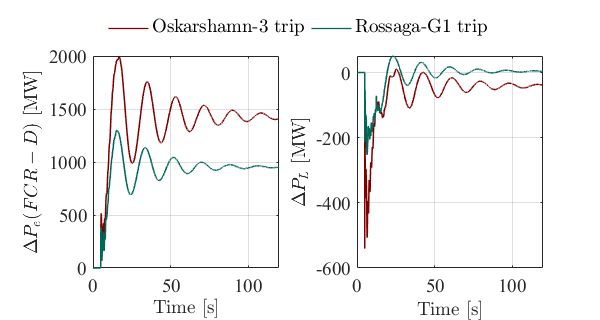}
\caption{Cumulated variations of electrical powers from FCR-D units (left plot) and load active powers (right plot)}\label{fig:v_load}
\end{figure}

The loss of voltage support by the tripped generator impacts the nearby voltages and, hence, the load power, which in turns affects the frequency response. By way of illustration, Fig.~\ref{fig:v_load} shows the cumulated variations of, respectively, electrical powers from FCR-D units and load active powers. The oscillations in load active power result from the coupling between voltage magnitude and frequency: the voltage swings with the system frequency and the load power follows through its voltage dependency.

The outage of Oskarshamn-3 results in a voltage dip which at its maximum reduces the load consumption by 500~MW, while following the Rossaga-G1 incident, the maximum load decrease is around 250~MW. This significant difference is explained by the fact that the former incident involves a higher generation loss and impacts a higher load in its neighbourhood.

The final load contribution is negligible for the \mbox{Rossaga-G1} outage, as confirmed by the right plot in Fig.~\ref{fig:v_load}, because the load voltages almost recover their pre-disturbance values. In contrast, the permanent contribution of loads is higher after the Oskarshamn-3 outage. Indeed, voltages do not recover their pre-fault values and load contributes with 50~MW to the final frequency deviation, while FCR-D contributes with 1400~MW.

\subsection{NPS frequency dynamics with EPC} \label{ch:res_EPC}

If the current frequency control requirements are kept unchanged in the time horizon of this study, the FCR-D units will be unable to contain the IFD within $\pm$~1~Hz after dimensioning incident trip and could even breach the load shedding activation limit, as shown by Fig.~\ref{fig:f_system}. The results of this section show that HVDC EPC can efficiently contribute to correcting this situation and are compared to Base Case corresponding to the scenario described in Section \ref{ch:res_Fnps}, where Oskarsham-3 tripped.

\subsubsection{Case 1}

\begin{figure}[h!]
\includegraphics[width=.52\textwidth]{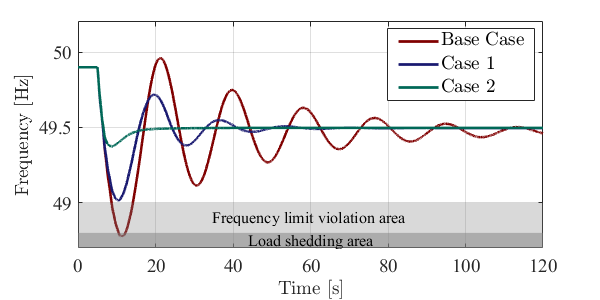}
\caption{NPS frequency response without and with HVDC EPC (complementing or replacing part of FCR-D)}\label{fig:EPC_f}
\end{figure}

\begin{figure}[h!]
\includegraphics[width=.52\textwidth]{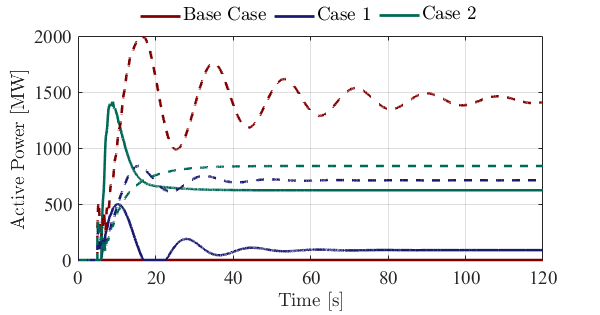}
\caption{Cumulated active power response of HVDC links (solid line) and FCR-D units (dashed lines) in the NPS}\label{fig:EPC_P}
\end{figure}

First, the case where HVDC EPC complements the FCR-D of generators is considered. Originally, ten generators provide FCR-D with a system stiffness $\beta_g=$~3648~MW/Hz. It was found that setting the frequency droops of HVDC EPC to a value as low as 0.33~pu is enough to meet the requirements in terms of nadir and final deviation of frequency. The corresponding contribution to system stiffness is $\beta_h =$ 519~MW/Hz. The resulting frequency evolution is shown in Fig.~\ref{fig:EPC_f}. From Eq.~(\ref{eq:Fss}) it is easily seen that, when the frequency settles below the triggering threshold $f_{TFL}$, HVDC links contribute to active power generation, which decreases the final frequency deviation. In this specific case, Fig.~\ref{fig:EPC_P} shows a cumulated contribution from HVDC links of 83~MW while frequency settles above 49.5~Hz (see Fig.~\ref{fig:EPC_f}).

Furthermore, Figs.~\ref{fig:EPC_f} and \ref{fig:EPC_P} clearly show that the fast correction applied by HVDC EPC significantly improves the damping of the common mode oscillation, compared to the situation with FCR-D from generators only.

\subsubsection{Case 2}

Next, the case where HVDC EPC partly replaces the FCR-D of generating units is considered. Thus FCR-D is procured by generators in areas asynchronous with NPS. To this purpose, three hydro generators in the NPS were removed from the FCR-D, which reduces their contribution to system stiffness $\beta_g$ by 1230~MW/Hz (changing from 3648 to 2418~MW/Hz). The missing system stiffness is compensated by HVDC EPC, re-tuned to achieve the same final frequency deviation of 0.5~Hz after the dimensioning incident. The new frequency droop is 0.0465~pu. This leads to increase the HVDC EPC contribution to system stiffness $\beta_h$ by 3196~MW/Hz (changing from 519 to 3715~MW/Hz). Thus, the increase of $\beta_h$ is 2.6 times larger than the decrease of $\beta_g$ that had to be compensated. This comes from the different frequency threshold used in DC/AC converters (49.6 Hz) with respect to that used in speed governors (49.9 Hz). 

With the above settings, the frequency nadir is raised to 49.36~Hz, as shown in Fig.~\ref{fig:EPC_f}. Thus, the NPS system operates with a comfortable security margin. As desired, the frequency settles at the final value of 49.5 Hz. 

It should be noted that HVDC links become dominant in the IFD containment, owing to the faster response of DC/AC converters (for similar values of the frequency droops). For instance, Fig.~\ref{fig:EPC_P} shows that, at the time of frequency nadir ($t=$~8.91~s), HVDC links inject a total 1,411~MW and generators in FCR-D only 697~MW. The frequency oscillation is also much better damped compared to the previous cases.  

The reverse of the medal is that the 1,411~MW increase in HVDC power flows account for 20\% of the available capacity at the considered operating point. In contrast, the frequency droop of 0.33~pu (see Case 1) imroves the IFD while using only 497.6~MW, accounting for only 7\% of the available capacity. The same tendency is noted in steady-state operation, with respectively 622~MW and 83~MW delivered to NPS. Although technically possible, FCR-D replacement with HVDC EPC requires substantial amounts of free capacity on the HVDC links. Presently several HVDC links connected to the NPS have more than 100~MW free capacity 70\% of the time on a yearly average, but the available capacity might decrease with larger power exchanges between countries, which would reduce the effectiveness of EPC \cite{nordpool} \cite{RB}. Measures such as capacity reservation, could resolve the issue at an additional price, but this must be coordinated with frequency reserve procurement in other areas. Currently, capacity reservation and frequency reserve procurement outside NPS is on a bilateral contract basis and coordinating and delivering the power reliably to NPS could be an issue.

\subsection{NPS frequency dynamics with support from NSWPH}

This section deals with NPS frequency support from the NSWPH, as well as frequency control in the NSWPH itself.

The control of the NSWPH frequency has been outlined in Section~\ref{sec:NSWPH}. It relies on the active power control loops of the offshore AC/DC terminal converters. In the low-inertia option the latter operate in grid-following mode with frequency droops set to 3.5 pu, a value that allows the SCs to play their role of kinetic energy buffers while preventing too large deviations of their rotor speeds (and, hence, of frequency). In the zero-inertia option, the same converters operate in grid-forming mode with frequency droops set to 0.01 pu to ensure small-signal stability of the offshore system \cite{SC}. In both cases, the integral gain $K_{hc}$ of the hub controller was set to 1.65 pu/s. The onshore AC/DC terminal converters control the DC voltages of their respective links.

HVDC EPC has been assigned to the link between NSWPH and Norway (see Fig.~\ref{fig:offshore_topology}), referred to as NSWPH-NO. The active power control loop of its offshore terminal is provided with a supplementary control receiving the onshore frequency error signal. The corresponding frequency droop was set to 0.33 pu. The same NPS dimensioning incident as in Sections~\ref{ch:res_Fnps} and \ref{ch:res_EPC} is considered. The simulation results correspond to the low-inertia option.

NSWPH is a system extension and, as such, it can be the source of additional disturbances. In this context, simulation results are provided pertaining to the outage of the HVDC link connecting NSWPH and The Netherlands (see Fig.~\ref{fig:offshore_topology}), referred to as NSWPH-NL. With a pre-disturbance power flow of 1743~MW, the loss of that link significantly impacts the active power balance of the isolated, offshore AC system. For comparison purposes, results are provided for the zero- and low-inertia designs of NSWPH.

\subsubsection{Dimensioning incident in NPS}

\begin{figure}[t!]
\hspace*{-4ex}
\includegraphics[width=.555\textwidth]{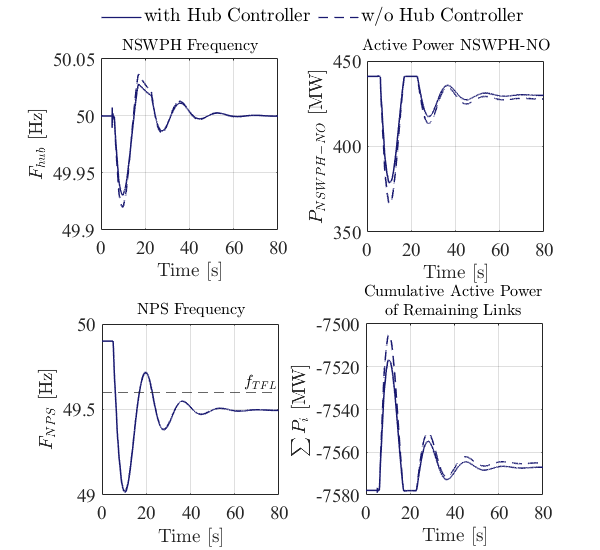}
\caption{Dimensioning incident in NPS with HVDC EPC on NSWPH-NO link and low-inertia option for NSWPH: evolution of frequencies and active power flows}
\label{fig:NSWPH_EPC}
\end{figure}

Fig.~\ref{fig:NSWPH_EPC} shows the NSWPH and NPS frequencies, the power flow in the \mbox{NSWPH-NO} link, and the sum of power flows in the other links connected to the NSWPH, in response to the dimensioning incident. At 6.02~s the NPS frequency drops below 49.6~Hz, thus activating EPC on the NSWPH-NO link. The power flow from NO to NSWPH is decreased to support the NPS frequency. 

The frequencies of the NSWPH and NPS systems are coupled through the EPC on the NSWPH-NO link, and the oscillations of the NSWPH frequency follow those of the NPS, through the power changes imposed by the terminal converter of that link. Furthermore, when the NPS frequency transiently exceed the $f_{FTL}$ threshold, shown with dashed line in the lower left plot in Fig.~\ref{fig:NSWPH_EPC}, the EPC contribution vanishes and the power flow in the NSWPH-NO link regains its initial value. A corresponding change of slope is observed in the evolution of the NSWPH frequency.

For an EPC droop value of 0.33~pu the NSWPH frequency deviates by less than 0.1~Hz. Clearly, larger deviations would be observed for larger values of the droop.

If EPC alone was considered, the power in the NSWPH-NO link would decrease to 367~MW at the time of NPS frequency nadir and would eventually settle to 427 MW. However, this is not the case since the offshore terminal converter of that link also participates in the control of the NSWPH frequency, as explained in Section~\ref{sec:NSWPH}. Since initially the NSWPH frequency decreases, the power flow from NO to NSWPH is increased, which slightly counteracts the decrease imposed by EPC. The simulation results show indeed a power flow of 379~MW at the time of NPS frequency nadir and 430~MW in the final steady state.

\begin{figure*}[h]
\centering
\includegraphics[width=1\textwidth]{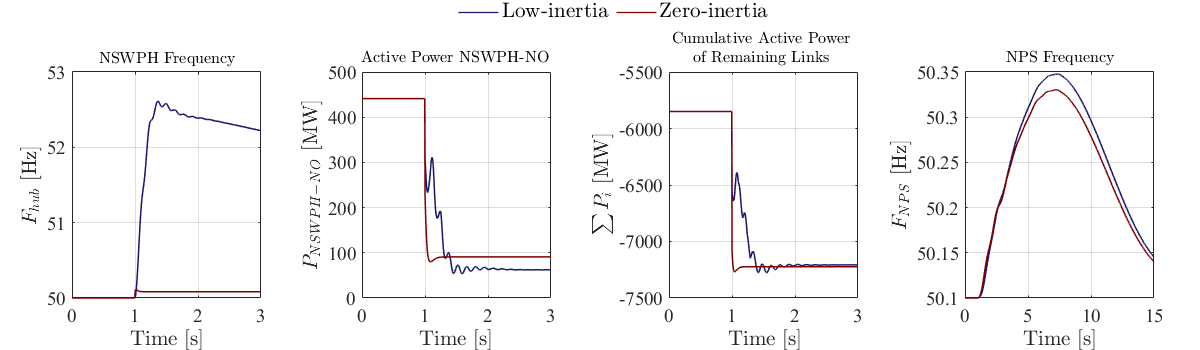}
\caption{Outage of the NSWPH-NL link: NSWPH and NPS frequencies, power flows in NSWPH-NO and sum of power flows in the other links, with the low- and zero-inertia designs}
\label{fig:NSWPH_comp}
\end{figure*}

\subsubsection{Outage of the NSWPH-NL HVDC link} 

This disturbance affects the active power balance of the NSWPH to a much greater extent. Figure~\ref{fig:NSWPH_comp} compares the low- and zero-inertia options in terms of NSWPH and NPS frequencies, power flow in the NSWPH-NO link, and sum of power flows in the other links connected to the NSWPH.

In the low-inertia design, the kinetic energy storage in the SCs yields a smoother active power adjustment of the offshore AC/DC converters. The rotor speeds and, hence, the NSWPH frequency, deviate quite significantly, with a maximum of 2.61~Hz. Although this deviation should be carefully taken into account when setting protections, no load is impacted in the offshore grid. Furthermore, the deviation could be made smaller by increasing the inertia of the rotating masses (using flywheels). The shown results correspond to a low value of 2 s for the inertia constant $H$ of the SCs. The frequency deviation is slowly corrected by the hub controller with integral action, bringing the NSWPH frequency  back to 50~Hz in some 80~s.

In contrast, in the absence of offshore energy storage, the zero-inertia design results in an almost instantaneous response of the AC/DC converters operating in grid-forming mode. The NSWPH frequency reaches its post-disturbance value in a few milliseconds and deviates from the nominal value by 0.12~Hz only. This is to be expected from the respective frequency droop values: 3.5~pu for the grid-following converters vs. 100~pu for the grid-forming ones. The small deviation will also be corrected by the integral hub controller. 

The almost instantaneous response of the zero-inertia design is also easily seen from the HVDC link powers. They reach their post-disturbance values in some 0.2~s with a tiny overshoot. The rate of change is less abrupt in the low-inertia case, resulting in a slower propagation of the disturbance in the HVDC links and to the onshore grids.

The steady-state difference observed for the power flow in the the NSWPH-NO link is due to the AC/DC converter of the NSWPH-GB link reaching its maximum current in the low-inertia scenario. This increases the participation in NSWPH frequency control of the other, non-limited AC/DC converters. By taking that limit into account in a more refined version of the hub controller, this discrepancy could be eliminated.
   
That change of power flow in the NSWPH-NO link disturbs the power balance of the NPS, in which it is felt as a load decrease. This causes a raise of the NPS frequency. The maximum IFD, reached at 7.3~s, amounts to 0.35~Hz in the low-inertia option, and 0.33~Hz in the zero-inertia option. The former deviation is slightly larger since the power flow in NSWPH-NO further decreases in the low-inertia scenario. The rate of change of frequency does not show a substantial difference, and the onshore frequency response does not significantly depend on the choice between zero- and low-inertia scenarios, for this specific disturbance.

\section{Conclusion}
In this paper, the Northern European AC/DC power system model is presented for a future low inertia NPS. Simulation results are reported for large disturbance with and without EPC, respectively. Moreover, a plausible NSWPH solution based on an AC collection hub is presented for low- and \mbox{zero-inertia} topologies enabling large wind power integration and additional frequency support to NPS.

It is shown that if the current frequency control requirements are kept unchanged and the dimensioning incident occurs, then it may not be possible to contain the frequency within the acceptable frequency range. This could trigger more severe frequency control mechanisms, such as load shedding, to counteract system degradation. It also shows that FCR-D alone may be unsuitable to manage the risk and that additional reserves may need to be procured from elsewhere. For example, IFD requirements could be satisfied by activating EPC on HVDC links. The simulation results show that HVDC EPC can be used to complement the existing FCR-D and can also be used to partly replace FCR-D, which was provided by conventional generating units. It was shown that frequency droop based HVDC EPC can considerably improve the IFD, due to the faster active power control of HVDCs compared to hydro generation.

When investigating the interaction between the NSWPH and the NPS, it was shown that the model can be used to study to which extend the hub can support the NPS and how disturbances on the hub propagate to the NPS. For that purpose, the model includes a low- and a zero-inertia configurations of the NSWPH.

\appendix[Share of frequency control between FCR-D and EPC]

It is considered that a fraction of FCR-D is procured outside the synchronous zone (namely by GB and CE in the case of NPS). The frequency containment process is characterized by its fast dynamic response, intended to limit the IFD, and by its participation in the final steady-state frequency deviation. The latter aspect is of concern hereafter. 

After a power disturbance $\Delta P_{dis}$, the system active power balance in steady state can be expressed as: 
\begin{align}
\displaystyle\sum_{i=1}^{g} P_{gi}^o &- (f -f_{FCR-D})\beta_g + \nonumber \\
+\displaystyle\sum_{i=1}^{l} &P_{hi}^o - (f -f_{TFL})\beta_h = P_L + \Delta P_{dis} \label{eq:PowerBalance}
\end{align}
where system stiffness provided from generators and HVDCs respectively are expressed as follows:
\begin{align}
\beta_g &=\frac{1}{f_N}\displaystyle\sum_{i=1}^{g}\frac{P_{Ngi}}{R_{gi}}   &   \beta_h &=\frac{1}{f_N}\displaystyle\sum_{i=1}^{l}\frac{P_{Nhi}}{R_{hi}}
\end{align}
and $g$ is the number of generators, $l$ the number of HVDC links contributing to EPC, $P_{gi}^o$ (resp. $P_{hi}^o$) the power setpoint of the $i$-th generator (resp. HVDC link), $P_{Ngi}$ (resp. $P_{Nhi}$) the nominal active power of the $i$-th generator (resp. HVDC link), $f$ (resp. $f_N$) the post-disturbance  steady-state (resp. nominal) frequency, $f_{FCR-D}$ (resp. $f_{TFL}$) the frequency threshold for activation of FCR-D (resp. EPC), $R_{gi}$ the steady-state speed droop of the $i$-th generator, $R_{hi}$ the frequency droop of the $i$-th HVDC link involved in EPC, $P_L$ the total active power consumption (including network losses).

Note that (\ref{eq:PowerBalance}) holds true only when the system frequency settles below $f_{TFL}$, enabling a permanent contribution from EPC in the post-fault configuration. Assuming that the system initially operates at the nominal frequency ($f=f_N$) the pre-disturbance power balance can be written as:
\begin{equation}\label{eq:predis}
\displaystyle\sum_{i=1}^{g} P_{gi}^o + \displaystyle\sum_{i=1}^{l} P_{hi}^o = P_L
\end{equation}
and (\ref{eq:PowerBalance}) can be rearranged to obtain the steady-state frequency deviation: 

\begin{align}
\label{eq:Fss}
\Delta f &= f-f_N = -\frac{1}{\beta_{g}+\beta_{h}} \Bigg[ \Delta P_{dist} + \nonumber \\
&+ (f_N-f_{FCR-D})\beta_g + (f_N-f_{TFL})\beta_h \Bigg]
\end{align}

(\ref{eq:PowerBalance}) shows that, for the same frequency droop values, HVDC links contribute less than generators, due to different frequency error input, owing to the lower activation threshold of EPC than for FCR-D. From (\ref{eq:Fss}) it can consistently checked that by replacing FCR-D with EPC without any adjustment of $R_{gi}$ and $R_{hi}$, the denominator term decreases while the bracket term increases. Both contribute to making $\Delta f$ more negative. Therefore, when replacing FCR-D with EPC, the HVDC link frequency droops must be properly selected to obtain the same steady-state frequency deviation.

\ifCLASSOPTIONcaptionsoff
  \newpage
\fi

\end{document}